# An alternative estimator for estimating the finite population mean in presence of measurement errors

Rajesh Singh, Sachin Malik

Department of Statistics, Banaras Hindu University

Varanasi-221005, India

**Absract**

This article presents the problem of estimating the population mean using auxiliary information in the presence of measurement errors. A numerical study is made among the proposed estimator, the exponential ratio estimator, Singh and Solanki (2012) estimator and the mean per unit estimator in the presence of measurement errors.

**Key words**: Population mean, Study variate, Auxiliary variates, Mean squared error, Measurement errors, Efficiency.

## 1. Introduction

In survey sampling, the properties of the estimators based on data usually presuppose that the observations are the correct measurements on characteristics being studied. Unfortunately, this ideal is not met in practice for a variety of reasons, such as non response errors, reporting errors, and computing errors. When the measurement errors are negligible small, the statistical inferences based on observed data continue to remain valid. on the contrary when they are not appreciably small and negligible, the inferences may not be simply invalid and inaccurate but may often lead to unexpected, undesirable and unfortunate consequences (See Srivastava and Shalabh,2001).Some authors including Allen et al.(2003), Manisha and Singh (2001.2002), Shalabh (1997) and Singh and Karpe (2008,2009) have paid their attention towards the estimation of population mean $\mu_y$ of the study variable y using auxiliary information in the presence of measurement errors.

For a simple random sampling scheme, let $(x_i, y_i)$ be observed values instead of the true values $(X_i, Y_i)$ on two characteristics (x, y) respectively for the $i^{th}$ (i=1.2….n) unit in the sample of size n.

Let the measurement errors be

$$u_i = y_i - Y_i \tag{1.1}$$

$$v_i = x_i - X_i \tag{1.2}$$

Which are stochastic in nature with mean zero and variances $\sigma_u^2$ and $\sigma_v^2$ respectively, and are independent. Further, let the population means of (x, y) be $(\mu_x, \mu_y)$, population variances of (x, y) be $(\sigma_x^2, \sigma_y^2)$ and $\sigma_{xy}$ and $\rho$ be the population covariance and the population correlation coefficient between x and y respectively. (See Manisha and Singh (2002)).

In this chapter we have studied the behaviour of some estimators in presence of measurement error.

## 2. Exponential ratio–type estimator under measurement error

Bahl and Tuteja (1991), suggested an exponential ratio type estimator for estimating $\bar{y}$ as

$$t_1 = \bar{y} \exp\left(\frac{\mu_x - \bar{x}}{\mu_x + \bar{x}}\right) \tag{2.1}$$

Let $w_u = \frac{1}{\sqrt{n}} \sum_{i=1}^{n} u_i$, $w_y = \frac{1}{\sqrt{n}} \sum_{i=1}^{n} (y_i - \mu_y)$

$w_v = \frac{1}{\sqrt{n}} \sum_{i=1}^{n} v_i$, $w_x = \frac{1}{\sqrt{n}} \sum_{i=1}^{n} (x_i - \mu_x)$

$C_x = \frac{\sigma_x}{\mu_x}$ and $C_y = \frac{\sigma_y}{\mu_y}$

Expression (2.1) can be written as

$$t_1 = \left[\mu_y + (\bar{y} - \mu_y)\right] \exp\left[\frac{\mu_x - (\mu_x + \bar{x} - \mu_x)}{\mu_x + (\mu_x + \bar{x} - \mu_x)}\right]$$

$$= (\mu_y + k_1)\exp\left[\frac{-k_2}{2\mu_x + k_2}\right] \tag{2.2}$$

where,

$$k_1 = \bar{y} - \mu_y = \frac{1}{\sqrt{n}}(w_y - w_u)$$

$$\text{and } k_2 = \bar{x} - \mu_x = \frac{1}{\sqrt{n}}(w_x - w_v), \text{var}(\bar{y}) = \frac{\sigma_y^2}{n}\left[1 + \frac{\sigma_u^2}{\sigma_y^2}\right]$$

On simplifying expression (2.2), we have

$$(t_1 - \mu_y) = \mu_y\left[-\frac{1}{2}\left(\frac{k_2}{\mu_x}\right) + \frac{3}{8}\left(\frac{k_2}{\mu_x}\right)^2 + ..\right] + k_1\left[1 - \frac{1}{2}\left(\frac{k_2}{\mu_x}\right) + \frac{3}{8}\left(\frac{k_2}{\mu_x}\right)^2 + ..\right] \tag{2.3}$$

On taking the expectations and using the results as

$$E(k_1) = E(k_2) = 0$$

$$E(k_1^2) = \frac{\sigma_y^2}{n}\left(1 + \frac{\sigma_u^2}{\sigma_y^2}\right) = V_{ym}$$

$$E(k_2^2) = \frac{\sigma_x^2}{n}\left(1 + \frac{\sigma_v^2}{\sigma_x^2}\right) = V_{xm}$$

$$E(k_1 k_2) = \frac{\rho \sigma_y \sigma_x}{n} = V_{yxm}$$

Taking expectation on both side of (2.3) the bias of $t_1$, to first order of approximation is

$$\text{Bias}(t_1) = \frac{1}{\mu_x}\left(\frac{3}{8}R_m V_{xm} - \frac{1}{2}V_{yxm}\right) \tag{2.4}$$

where $R_m = \dfrac{\mu_y}{\mu_x}$

Squaring of both side of (7.2.3) and taking expectation, the mean square error of $t_1$ up to first order of approximation, is

$$MSE(t_1) = E(t_1 - \mu_y)^2$$

$$= \frac{\mu_y^2}{4\mu_x^2} E(k_2^2) + E(k_1^2) - \frac{\mu_y}{\mu_x} E(k_1 k_2)$$

$$= \frac{1}{4} R_m^2 \left( \frac{\sigma_x^2}{n} \left( 1 + \frac{\sigma_v^2}{\sigma_x^2} \right) \right) + \frac{\sigma_y^2}{n} \left( 1 + \frac{\sigma_u^2}{\sigma_y^2} \right) - \frac{\rho \sigma_y \sigma_x}{n} R_m$$

$$MSE(t_1) = \frac{\sigma_y^2}{n} \left[ 1 - \frac{C_x}{C_y} \left( \rho - \frac{C_x}{4C_y} \right) \right] + \frac{1}{n} \left[ \frac{\mu_y^2}{4\mu_x^2} \sigma_v^2 + \sigma_u^2 \right] \quad (2.5)$$

$$MSE(t_1) = M_{t_1}^* + M_{t_1}$$

where,

$$M_{t_1}^* = \frac{\sigma_y^2}{n} \left[ 1 - \frac{C_x}{C_y} \left( \rho - \frac{C_x}{4C_y} \right) \right]$$ is the mean squared error of $t_1$ without measurement error.

and

$$M_{t_1} = \frac{1}{n} \left[ \frac{\mu_y^2}{4\mu_x^2} \sigma_v^2 + \sigma_u^2 \right]$$ Is the contribution of measurement errors in $t_1$.

## 3. Another estimator under measurement error

Koyuncu and Kadilar (2010), suggested a regression type estimator $t_2$ as-

$$t_2 = \omega_1 \bar{y} + \omega_2 (\mu_x - \bar{x}) \quad (3.1)$$

where $\omega_1$ and $\omega_2$ are constants that have no restriction.

Expression (3.1) can be written as

$$t_2 - \mu_y = \omega_1 k_1 + \mu_y (\omega_1 - 1) - \omega_2 k_2 \quad (3.2)$$

Taking both side expectation of (3.2), we get the bias of the estimator $t_2$ to order $O(n^{-1})$

$$\text{Bias}(t_2) = \mu_y(\omega_1 - 1)$$

Squaring both side of (3.2) and taking expectation, the MSE of $t_2$ to the to order $O(n^{-1})$ is

$$\text{MSE}(t_2) = \mu_y^2(\omega_1 - 1)^2 + \frac{1}{n}\omega_1^2\sigma_y^2 + \frac{1}{n}\omega_2^2\sigma_x^2 - \frac{2}{n}\omega_1\omega_2\rho\sigma_y\sigma_x + \frac{1}{n}\left(\omega_1^2\sigma_u^2 + \omega_2^2\sigma_v^2\right) \quad (7.2.14)$$

$$= M_{t_2}^* + M_{t_2} \quad (3.3)$$

where,

$$M_{t_2}^* = \mu_y^2(\omega_1 - 1)^2 + \frac{1}{n}\omega_1^2\sigma_y^2 + \frac{1}{n}\omega_2^2\sigma_x^2 - \frac{2}{n}\omega_1\omega_2\rho\sigma_y\sigma_x, \text{ is the MSE of } t_2 \text{ without measurement error and}$$

$$M_{t_2} = \frac{1}{n}\left(\omega_1^2\sigma_u^2 + \omega_2^2\sigma_v^2\right), \text{ is the contribution of measurement error in } t_2.$$

Writing MSE of $t_2$ as

$$\text{MSE}(t_2) = (\omega_1 - 1)^2\mu_y^2 + \omega_1^2(a_1) + \omega_2^2 a_2 + 2\omega_1\omega_2(-a_3) \quad (3.4)$$

where,

$$a_1 = (V_{ym}), \qquad a_2 = (V_{xm}), \quad \text{and} \quad a_3 = (V_{yxm})$$

Now, optimising MSE of the estimator $t_2$ with respect to $\omega_1$ and $\omega_2$, we get

$$\omega_1^* = \frac{b_3 b_4}{b_1 b_3 - b_2^2} \quad \text{and} \quad \omega_2^* = -\frac{b_2 b_4}{b_1 b_3 - b_2^2} \quad (3.5)$$

where,

$$b_1 = \mu_y^2 + a_1, \qquad b_2 = -a_3, \qquad b_3 = a_2, \quad \text{and} \quad b_4 = \mu_y^2.$$

Using the values of $\omega_1^*$ and $\omega_2^*$ from equation (3.5) into equation (3.4) we get the minimum MSE of the estimator $t_2$ as

$$\text{MSE}(t_2)_{min} = \left[\mu_y^2 - \frac{b_3 b_4^2}{b_1 b_3 - b_2^2}\right] \tag{3.6}$$

Singh and Solanki suggested an estimator $t_3$ as

$$t_3 = \bar{y}\left\{2 - \left(\frac{\bar{x}}{\mu_x}\right)^\alpha \exp\left[\frac{\beta(\bar{x} - \mu_x)}{(\bar{x} + \mu_x)}\right]\right\} \tag{3.7}$$

Expanding equation (3.7) and subtracting $\mu_y$ from both side

$$(t_3 - \mu_y) = -\mu_y\left[-\frac{V_{xm}}{\mu_x^2}\left(\alpha(\alpha-1) + \frac{\beta(\beta-2)}{8} + \frac{\alpha\beta}{2}\right)\right] - \frac{V_{yxm}}{\mu_x}\left(\alpha + \frac{\beta}{2}\right) \tag{3.8}$$

On taking expectation of both side of (3.8) we get the bias of the estimator $t_3$ to the order $O(n^{-1})$

$$\text{Bias}(t_3) = \mu_y\left\{\frac{V_{xm} A}{\mu_x^2}\right\} - \left\{\frac{B}{\mu_x} V_{xm}\right\} \tag{3.9}$$

where,

$$A = \left[\alpha(\alpha-1) + \frac{\beta(\beta-2)}{8} + \frac{\alpha\beta}{2}\right], \quad \text{and} \quad B = \left(\alpha + \frac{\beta}{2}\right)$$

Squaring both side of (3.8) and taking expectations, the MSE of $t_3$ to the order $O(n^{-1})$ is

$$\text{MSE}(t_3) = E(t_3 - \mu_y)^2$$
$$= V_{yxm} + V_{xm} R_m^2 B^2 - 2R_m V_{yxm} B \tag{3.10}$$

## 4. A general class of estimators

Following Singh and Solanki, we propose a general class of estimator $t_3$ as

$$t_4 = [m_1 \bar{y} + m_2(\mu_x - \bar{x})]\left\{2 - \left(\frac{\bar{x}}{\mu_x}\right)^\alpha \exp\left[\frac{\beta(\bar{x} - \mu_x)}{(\bar{x} + \mu_x)}\right]\right\} \tag{4.1}$$

Expanding equation (4.1) and subtracting $\mu_y$ from both side

$$(t_4 - \mu_y) = \left\{(m_1 - 1)\mu_y - m_1\mu_y\left\{B\frac{k_2}{\mu_x} + \frac{k_2^2 A}{\mu_x^2}\right\} + m_1 k_1\left\{1 - \frac{Bk_2}{\mu_x}\right\} - m_2 k_2\left\{1 - \frac{Bk_2}{\mu_x}\right\}\right\} \quad (4.2)$$

On taking expectation of both side of (4.2) we get the bias of the estimator $t_4$ to the order $O(n^{-1})$

$$\text{Bias}(t_4) = (m_1 - 1)\mu_y - m_1\mu_y\left\{\frac{V_{xm} A}{\mu_x^2}\right\} - m_1\left\{\frac{B}{\mu_x} V_{yxm}\right\} + m_2\left\{\frac{B}{\mu_x} V_{xm}\right\} \quad (4.3)$$

Squaring both side of (4.2) and taking expectations, the MSE of $t_4$ to the order $O(n^{-1})$ is

$$\text{MSE}(t_4) = E(t_4 - \mu_y)^2$$

$$= (m_1 - 1)^2 \mu_y^2 + m_1^2 R_m^2 B^2 V_{xm} + m_1^2 V_{yxm} - 2m_1^2 R_m B V_{yxm} - 2m_1(m_1 - 1) R_m^2 A V_{xm} + m_2^2 V_{xm}$$

$$+ 2\{m_2(m_1 - 1)BR_m V_{xm} + m_1 m_2 R_m V_{xm} - m_1 m_2 V_{yxm} + m_1(m_1 - 1)BR_m V_{yxm}\}$$

The MSE of the estimator $t_3$ can also be written as

$$\text{MSE}(t_4) = (m_1 - 1)^2 \mu_y^2 + m_1^2 P_1 + m_2^2 P_2 + 2m_1 m_2 P_3 - 2m_1 P_3 - 2m_2 PA_5 \quad (4.4)$$

where,

$P_1 = (V_{ym} + B^2 R_m^2 V_{xm} - 2R_m^2 A V_{xm})$, $\quad P_2 = (V_{xm})$

$P_3 = (2BR_m V_{xm} - V_{yxm})$, $\quad P_4 = (R_m B V_{yxm} - A V_{xm} R_m^2)$

$P_5 = (BR_m V_{xm})$

Now optimising MSE $t_3$ with respect to, $m_1$ and $m_2$, we get the optimum values -

$$m_1^* = \frac{B_1 P_2 - P_3 B_5}{B_2 P_2 - P_3^2} \text{ and } m_2^* = \frac{B_2 P_5 - B_1 P_3}{B_2 P_2 - P_3^2}$$

where,

$$B_1 = \mu_y^2 + P_4 \qquad B_2 = \mu_y^2 + P_1$$

## 5. Theoretical Efficiency Comparisons:

The MSE of the proposed estimator $t_1$ proposed in (2.1) will be smaller than usual estimator under measurement error case if the following condition is satisfied by the data set

$$\frac{\sigma_y^2}{n}\left[1-\frac{C_x}{C_y}\left(\rho-\frac{C_x}{4C_y}\right)\right]+\frac{1}{n}\left[\frac{\mu_y^2}{4\mu_x^2}\sigma_v^2+\sigma_u^2\right] \leq \frac{\sigma_y^2}{n}\left(1+\frac{\sigma_u^2}{\sigma_y^2}\right)$$

or

$$R_m^2 \frac{V_{xm}}{V_{yxm}} \leq 4 \qquad (5.1)$$

As we know that the estimators $t_2$ defined in (3.1) is the particular members of the generalised estimator $t_3$ so, if the above condition is satisfied for different values of $\alpha, \beta$, and $m_1, m_2$ the estimator $t_3$ or $t_2$ will be better than usual estimator under measurement errors.

$$MSE(t_4)_{min} \leq V(\bar{y}_m) \qquad (5.2)$$

## 6. Empirical Studies:

**Data statistics:** The data used for empirical study has been taken from Gujrati and Sangeetha (2007) -pg, 539.

Where, $Y_i$= True consumption expenditure,

$X_i$= True income,

$y_i$= Measured consumption expenditure,

$x_i$ = Measured income.

From the data given we get the following parameter values

**Table.1:**

| n | $\mu_y$ | $\mu_y$ | $\sigma_y^2$ | $\sigma_x^2$ | $\rho$ | $\sigma_u^2$ | $\sigma_v^2$ |
|---|---|---|---|---|---|---|---|
| 10 | 127 | 170 | 1278 | 3300 | 0.964 | 36.00 | 36.00 |

-

| Estimators | Values of $\alpha$ and $\beta$ | PRE |
|---|---|---|
| $\bar{y}_m$ | - | 100.00 |
| $t_1$ | - | 437.59 |
| $t_{regm}$ | $\alpha = 1, \beta_m = \dfrac{V_{yxm}}{\sqrt{V_{xm}V_{ym}}}$ | 946.54 |
| $t_{2min}$ | $\alpha = 0, \beta = 0$ | 944.94 |
| $t_{3min}$ | $\alpha = 1, \beta = 1$ <br> $\alpha = 1, \beta = 0$ <br> $\alpha = 0, \beta = 1$ <br> $\alpha = 1, \beta = -1$ | 123.23 <br> 603.01 <br> 437.27 <br> 437.27 |
| $t_{4min}$ | $\alpha = 1, \beta = 0$ | 1012.77 |
| | $\alpha = 0, \beta = 1$ | 1031.13 |
| | $\alpha = 1, \beta = 1$ <br> $\alpha = 1, \beta = -1$ | 948.35 <br> **1031.11** |

**Table (5.3): Showing the MSE of the estimators with and without measurement errors**

| Estimators | MSE without meas. Error | Contribution of meas. Error in MSE | MSE with meas. Error |
|---|---|---|---|
| $\bar{y}_m$ | 127.800 | 3.600 | 131.400 |
| $t_1$ | 25.925 | 4.102 | 30.028 |
| treg | 9.000 | 4.896 | 13.882 |
| $t_{2min}$ $(\alpha=0, \beta=0)$ | 8.995 | 4.910 | 13.905 |
| $t_{3min}$ $(\alpha=1, \beta=0)$ | 17.203 | 4.587 | 21.790 |
| $(\alpha=0, \beta=1)$ | 25.798 | 4.252 | 30.050 |
| $(\alpha=1, \beta=1)$ | 101.874 | 4.747 | 106.621 |
| $(\alpha=1, \beta=-1)$ | 25.772 | 4.278 | 30.050 |
| $t_{4min}$ $(\alpha=1, \beta=0)$ | 8.397 | 4.577 | 12.974 |
| $(\alpha=0, \beta=1)$ | 8.536 | 4.207 | 12.743 |
| $(\alpha=1, \beta=1)$ | 8.990 | 4.865 | 13.855 |
| $(\alpha=1, \beta=-1)$ | 7.868 | 4.874 | **12.742** |

**Conclusion:**

From the Table 5.3, we conclude that the effect due to measurement error on the exponential ratio type and other proposed estimators is less than the effect on the $t_2$ and regression estimators while it is more serious than usual estimator under measurement error for this data set.